\documentclass[prb,twocolumn,amsmath,amssymb,letterpaper,superscriptaddress,nofootinbib]{revtex4-2}
\usepackage{mathrsfs,braket}
\usepackage{amssymb, amsbsy, amsmath, latexsym, dsfont, array, layout, graphicx, mathrsfs, color, ulem, bm}
\usepackage[colorlinks,linkcolor=blue,filecolor=blue,urlcolor=blue,citecolor=blue]{hyperref}

\begin{document}
\title{Anomalous localization and duality in
non-Hermitian quasiperiodic models}

 \author{Wenzhi Wang}
 \affiliation{Laboratory of Quantum Information, University of Science and Technology of China, Hefei 230026, China}
 \author{Tianyu Li}
\email{tianyuli@m.scnu.edu.cn}
\affiliation {Key Laboratory of Atomic and Subatomic Structure and Quantum Control (Ministry of Education), Guangdong Basic Research Center of Excellence for Structure and Fundamental Interactions of Matter, School of Physics, South China Normal University, Guangzhou 510006, China}
\affiliation {Guangdong Provincial Key Laboratory of Quantum Engineering and Quantum Materials, Guangdong-Hong Kong Joint Laboratory of Quantum Matter, Frontier Research Institute for Physics, South China Normal University, Guangzhou  510006, China}
\author{Wei Yi}
\email{wyiz@ustc.edu.cn}
\affiliation{Laboratory of Quantum Information, University of Science and Technology of China, Hefei 230026, China}
\affiliation{Anhui Province Key Laboratory of Quantum Network, University of Science and Technology of China, Hefei 230026, China}
\affiliation{CAS Center For Excellence in Quantum Information and Quantum Physics, Hefei 230026, China}
\affiliation{Hefei National Laboratory, University of Science and Technology of China, Hefei 230088, China}

\begin{abstract}

We develop a general framework based on the sign patterns of Lyapunov exponents for engineering and characterizing counterintuitive localization properties in one-dimensional non-Hermitian quasiperiodic lattices.
On the one hand, for the Anderson localized states in the bulk under the periodic boundary condition, we find that their localization features can be boundary-sensitive, becoming boundary-localized skin modes under the open boundary condition.
On the other hand, for non-localized states, the well-known extended-localized duality relation can break down, as their counterparts in the dual model can also be nonlocal.
We discuss how these remarkable phenomena can be understood from the perspective of Lyapunov exponents, highlighting the interplay of non-Hermiticiy and quasiperiodicity.
\end{abstract}

\maketitle

\section{Introduction}
Anderson localization is a fundamental paradigm in physics, where wave propagations are localized by disorder~\cite{Anderson1958, Lee1985, Lagendijk2009, Abrahams1979, Kramer1993}. 
In quasiperiodic lattice models, Anderson localization also arises, often accompanied by the localization-delocalization transition~\cite{Harper1955, Kohmoto1983, Roati2008, Thouless1988, Deng2019, Wang2016, Szabo2020, Borgnia2022, Hetenyi2025}. 
Take the Aubry-Andr\'{e}-Harper model for instance~\cite{Aubry1980, Jitomirskaya1999, Modugno2009}. 
The critical point for the transition is identified through self-duality, where the model Fourier transforms into itself. 
Away from the self-dual critical point, an extended-localized duality relation persists, under which an extended state is mapped to a localized one in the dual model, and vice versa~\cite{Aubry1980, Biddle2010, Ganeshan2015}. 
Such a duality relation has been considered as a fundamental guiding principle for the study of quasiperiodic models~\cite{Gopalakrishnan2017, Goncalves2023, Sahu2021, Hu2025,lizhi2026}. 

Localization and self-duality also emerge in non-Hermitian quasiperiodic models~\cite{Longhi2019, Jiang2019,Cai2021, Cai2025}, where the phases and phase transitions are significantly enriched by unique features of non-Hermitian systems such as the spectral topology and non-Hermitian symmetries~\cite{Gong2018, Kawabata2019, Kawabata2020, Okuma2020, Bergholtz2021,Cai2021,Cai2025}. 
Like the Hermitian case, the extended-localized duality relation has facilitated the study of various non-Hermitian quasiperiodic models~\cite{Zeng2020a, Liu2020a, Liu2021,Liu2021b, Longhi2019b}. 
Further, despite the potential boundary sensitivity of eigenstates and eigenspectra in non-Hermitian systems, as in the case of the non-Hermitian skin effect~\cite{Yao2018, Song2019, Yokomizo2019, Zhang2020, Ashida2020, Zhang2022, Okuma2023, CHLee2023}, it has been considered a universal axiom that localized states in the bulk are insensitive to boundary conditions~\cite{Liu2021b,Liu2021,Sun2025}.

In this work, we develop a general Lyapunov-exponent perspective for characterizing localization properties in non-Hermitian quasiperiodic lattices. By classifying states according to the sign patterns of the Lyapunov exponents, we discuss the engineering of anomalous, boundary-sensitive localized states and the breakdown of the extended-localized duality relation in non-Hermitian quasiperiodic lattices.
Specifically, the boundary-sensitive localized states are Anderson localized in the bulk under the periodic boundary condition (PBC), but become boundary-localized skin modes under the open boundary condition (OBC). While such a boundary-sensitive localization was recently reported for a ladder system with rungs coupling a non-Hermitian Hatano--Nelson model and an Aubry--Andr\'{e}--Harper model~\cite{SZLi2025}, its origin and the general criteria for its emergence remains elusive. Here, we demonstrate that this counterintuitive localization phenomenon is governed by the sign patterns of the LEs. We further show that the conventional extended--localized duality can break down, such that the duals of extended states are still extended, provided they are subject to the non-Hermitian skin effect in both models. 
These counterintuitive properties originate from the interplay of quasiperiodicity and the non-Hermitian skin effect (more generally, the non-Hermitian boundary-sensitivity), whereas insight and control to these anomalous behaviors are provided by the Lyapunov exponents\cite{Sun2025, Avila2015, SZLi2025}.
Our findings shed new light on localization in non-Hermitian quasiperiodic systems.

The work is organized as follows. In Sec.~II, we discuss the general sign patterns of the Lyapunov exponents, and their impact on the localization properties under different boundary conditions. 
As concrete examples of its impact, we study the boundary-sensitive localization and the breakdown of duality in Sec.~III and Sec.~IV, respectively. A briefly summary is given in Sec.~V.

\section{Lyapunov exponents and localization}
We start by analyzing the general patterns of Lyapunov exponents for quasiperiodic models. 
Consider a one-dimensional tight-binding Hamiltonian $\hat{H}$ with $N$ lattice sites and hopping range
$M$:
\begin{equation}
\hat{H}=\sum_{|n-m|\leq M}J_{n,m}\hat{b}_{n}^{\dagger}\hat{b}_{m},\label{eq:H_Jnm}
\end{equation}
where $\hat{b}_{n}^{\dagger}$ ($\hat{b}_{n}$) denote the
creation (annihilation) operator on site $n$, $J_{n,m}$ is the
hopping amplitude from site $m$ to $n$ for $n\neq m$, or the on-site potential for $n=m$.
The corresponding eigenvalue equations read
\begin{equation}
\sum_{t=-M}^{M}J_{n,n+t}\psi_{n+t}=E\psi_{n}, \quad n = 1,2,...N,\label{eq:eigen_equa}
\end{equation}
where $\psi_{n}$ is the eigenstate amplitude on site $n$.
The $2M\times2M$ transfer matrix with eigenenergy $E$ is
\begin{equation}
T(E)=\prod_{n=1}^{N}T_{n}(E),\label{eq:trans_full}
\end{equation}
where $T_n$ is defined through $\Psi_{n}=T_{n}(E)\Psi_{n-1}$ and $\Psi_{n}=(\psi_{n+M},\psi_{n+M-1},\cdots,\psi_{n-M+1})^{T}$ (see Appendix~\ref{app:transfer}).
Here, to impose the PBC, we set $\hat{c}_{j+N} = \hat{c}_j$ and $\hat{c}_{j+N}^\dagger = \hat{c}_j^\dagger$. It follows that $\psi_{j}=\psi_{j-N}$ for $j>N$ and $\psi_{j}=\psi_{j+N}$ for $j\leq 0$. For the OBC, all hopping terms crossing the boundaries are removed, so that $\psi_j=0$ for $j>N$ or $j\leq 0$.

Importantly, the eigenvalues of the matrix $\frac{1}{2N}\ln[T^{\dagger}(E)T(E)]$ in the thermodynamic
limit $N\rightarrow+\infty$ correspond to the $2M$ Lyapunov exponents $\gamma_{i}(E)$ ($i=1,\cdots,2M$),
from which the localization properties of the model can be obtained~\cite{Hoffmann1996}. 
For convenience, we sort the Lyapunov exponents in ascending order, with $\gamma_{1}(E)\leq\cdots\leq\gamma_{2M}(E)$. 
With the prefactor $1/N$ in Eq.~(\ref{eq:trans_full}), the boundary contributions 
are negligible in the thermodynamic limit.
Hence, for a fixed eigenenergy $E$, one obtains the same set of Lyapunov exponents $\{\gamma_i(E)\}$ under different boundary conditions~\cite{Sun2025}. 
Nevertheless, as we show below, localization behaviors of the corresponding eigenstates can 
be distinct under different boundary conditions.

\begin{figure}[tbp]
\centering

\includegraphics[scale=0.55]{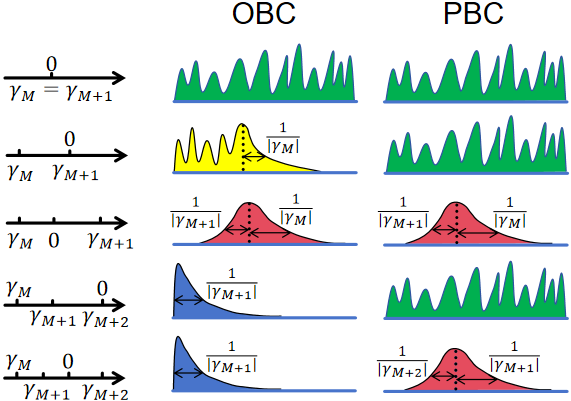}

\caption{Schematics of spatial profiles of eigenstates under the OBC and PBC for different sign patterns of 
the Lyapunov exponents. Green profiles denote non-localized states, red ones denote Anderson localized states, blue denotes boundary-localized skin modes, and yellow denotes states that are extended in one direction and localized in the other~\cite{Sun2025}. The quantity $1/|\gamma_i|$ indicates the corresponding localization length.\label{fig:1}}

\end{figure}

Let us first consider models without disorder or quasiperiodicity, and the case where the hopping amplitudes in the bulk possess the translational symmetry, with  $J_{n,m}=J_{n-m} = t_s$ and $s = n-m$. 
The corresponding Bloch Hamiltonian is then $H(k) = \sum_{s=-M}^{M} t_s e^{iks}$. 
A unified representation of the eigenstates and eigenspectrum, regardless of Hermiticity and boundary conditions, can be formally achieved through the non-Bloch band theory~\cite{Kawabata2020, Yao2018, Yokomizo2019}, where the factor $e^{ik}$ is replaced by a complex parameter $\beta$. For the Hermitian case or for non-Hermitian systems under the PBC, $\beta$ reduces to the Bloch form of $|\beta|=1$. But in general, the characteristic equation $\sum_{s=-M}^{M} t_s \beta^{s}-E=0$ admits $2M$ solutions 
$\beta_i(E)$ (with $i=1,\ldots,2M$ and labeled in ascending order of $|\beta_i|$). 
In the thermodynamic limit, eigenspectrum of the system under the OBC is given by eigenenergies $E$ that satisfy 
$|\beta_M(E)|=|\beta_{M+1}(E)|$. 
The corresponding eigenstates under the OBC are constructed from the solutions $\beta_M$ and $\beta_{M+1}$,
where the Bloch wave factor $e^{ikn}$ is replaced by $\beta_M^n$. As a result, localization toward one of the open boundaries, that is, the non-Hermitian skin effect, arises for $|\beta_M|\neq 1$. 
Crucially, the solutions of the characteristic equation $\beta_i$ are directly related to the Lyapunov exponents through $\gamma_i(E)=\ln|\beta_i(E)|$~\cite{Sun2025}. 
This means that, under the OBC, the Lyapunov exponents $\gamma_M=\gamma_{M+1}$ are solely responsible for the localization properties of the system.
When disorder or quasiperiodicity is introduced, $\gamma_{M}(E)$ and $\gamma_{M+1}(E)$ are generally no longer equal, but they still dominate the spatial distributions of eigenstates under the OBC~\cite{Sun2025}.

By contrast, under the PBC, the spatial distribution of eigenstates may no longer be dominated by $\gamma_{M}(E)$ and $\gamma_{M+1}(E)$. In general, different Lyapunov exponents $\gamma_{i}(E)$ correspond to different decay channels of the eigenstates with eigenenergy $E$. 
The channel associated with $\gamma_{i}(E)$ contribute a factor $e^{\gamma_{i}(E)n}$ to the spatial profile of the corresponding eigenstate. 
Hence, in the presence of a vanishing exponent $\gamma_i(E)=0$, the eigenstate is not Anderson localized, but 
either extended or critical~\cite{Avila2015, Liu2021b}.
Otherwise, in the case that all $\gamma_i(E)\neq 0$, eigenstates at the given $E$ must be localized, since channels with positive $\gamma_i$ (implying an exponential wave-function growth in $n$) and those with negative $\gamma_i$ (n exponential decay in $n$) must coexist. This is because the $2M$ Lyapunov exponents corresponding to a given
eigenenergy $E$ under the PBC cannot all have the same sign, otherwise the monotonic growth (decay) of the wave function would contradict the PBC.
Importantly, the localization length on the two sides of the localization center should be governed by the pair of Lyapunov exponents with opposite signs that are closest to zero.

In the conventional case, the Lyapunov exponents are distributed symmetrically with respect to zero. Hence, 
the dominant decay channels are still governed by $\gamma_M$ and $\gamma_{M+1}$, and the localization behaviors of a given eigenstate are identical under the PBC and OBC. 
However, by introducing nonreciprocity, the Lyapunov exponents $\gamma_i(E)$ acquire a shift (see Appendix~\ref{app:shift}), 
which can break the symmetry in the sign pattern of the Lyapunov exponents. 
As a result, the Lyapunov exponents governing the spatial distribution of eigenstates 
can be different under the PBC and OBC. 
As we show below, this can lead to scenarios beyond the standard non-Hermitian skin effect.

We list the five typical patterns of the Lyapunov exponents in Fig. 1, as well as the corresponding localization scenarios under the OBC and PBC, respectively:
(i) $\gamma_{M}=\gamma_{M+1}=0$, the eigenstates are extended or critical, regardless of the boundary condition;
(ii) $\gamma_{M}<\gamma_{M+1}=0$ (or $\gamma_{M}=0<\gamma_{M+1}$), 
the corresponding egienstates are extended or critical under the PBC, but is localized in one direction while nonlocal in the other~\cite{Sun2025};
(iii) $\gamma_{M}<0<\gamma_{M+1}$, the eigenstates are Anderson localized under both OBC and PBC;
(iv) $\gamma_{M}<\gamma_{M+1}<\gamma_{M+2}=0$ (or $\gamma_{M-1}=0<\gamma_{M}<\gamma_{M+1}$),
the eigenstates are extended or critical under the PBC, but boundary-localized under the OBC, as in the standard non-Hermitian skin effect;
(v) $\gamma_{M}<\gamma_{M+1}<0<\gamma_{M+2}$ (or $\gamma_{M-1}<0<\gamma_{M}<\gamma_{M+1}$)
, where the eigenstates are Anderson localized under the PBC, but boundary-localized under the OBC.

Note that some of the OBC cases above have already been discussed in Ref.~\cite{Sun2025}. Importantly, in contrast to Ref.~\cite{Sun2025} where the localization properties are determined solely by the central Lyapunov exponents $\gamma_M$ and $\gamma_{M+1}$, our analysis reveals that other Lyapunov exponents can also play an important role. In particular, exponents beyond the central pair, such as $\gamma_{M+2}$, may become relevant for characterizing the spatial structure of the eigenstates. This is exemplified by 
case (v), where the localization is boundary-sensitive. It is distinct from either the conventional Anderson localization, which is unaffected by the boundary, or the conventional non-Hermitian skin effect, where the states are extended or critical under the PBC~\cite{Liu2021b, Cai2022}.
Importantly, these boundary-sensitive localized states can be engineered by tuning the pattern of the Laypunov exponents.

\section{Boundary-sensitive localization}
As an illustrative example, we consider a one-dimensional non-Hermitian quasiperiodic model with both nearest-neighbor and next-nearest-neighbor hoppings ($M = 2$)
\begin{equation} \hat{H}_{1}(g,h)=\sum_{1\leq|m-n|\leq2}e^{(n-m)g}\hat{b}_{m}^{\dagger}\hat{b}_{n}\frac{\cos[\tau\pi(m+n)+ih]}{|m-n|^{3}}.\label{eq:H1} 
\end{equation}
Here the irrational number $\tau=(\sqrt{5}-1)/2$, and the non-Hermitian parameters $h$ and $g$ take real values.
The Hamiltonian is a non-Hermitian extension of the self-dual design in Ref.~\cite{wang2025}, which facilitates our following discussion of its dual model.
Notably, by varying $g$ and $h$, the Lyapunov exponents either acquire a constant shift or becomes wider in range~\cite{Sun2025,Liu2021b}, enabling the engineering of particular patterns of the Lyapunov exponents.
To characterize localization, we use the fractal dimension (FD) of a normalized eigenstate
\begin{equation}
\text{FD}=-\frac{\ln\sum_{n}|\psi_{n}|^{4}}{\ln N}.
\label{eq:FD}
\end{equation}
In the thermodynamic limit, the FD approaches $0$ for localized states and $1$ for extended states, while it takes intermediate values between $0$ and $1$ for critical states.

\begin{figure}
\centering

\includegraphics[scale=0.215]{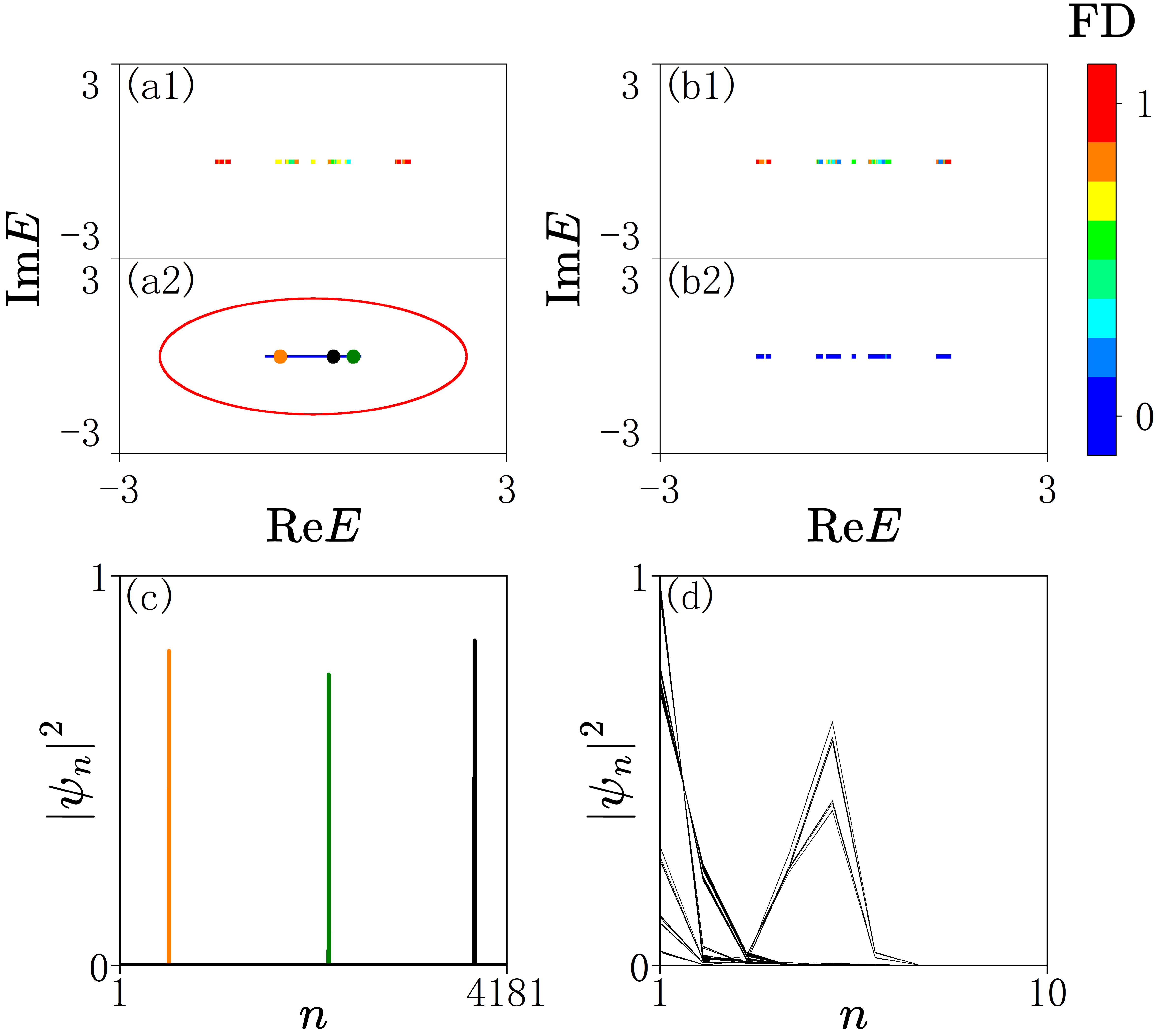}

\caption{(a)(b) Eigenspectra of $\hat{H}_{1}(g,0)$ under (a) PBC and (b) OBC, respectively,
where the color indicates the FD of the corresponding eigenstate. The parameters are (a1)(b1) $g=0$ and (a2)(b2) $g=1$. 
(c) Spatial distributions of typical localized states, marked in (a2) with matching colors.
(d) Spatial distributions of every tenth eigenstate in (b2). 
These eigenstates are all localized toward the boundary $n=1$.
The system size for all calculations is $N=4181$. \label{fig:2}}
\end{figure}

For convenience, we first set $h=0$. Then, if $g$ also vanishes, $\hat{H}_{1}$ becomes Hermitian, and its eigenstates consist of critical and extended states regardless of the boundary condition~\cite{wang2025}. 
This is confirmed in Fig.~\ref{fig:2}(a1)(b1), where the eigenspectrum remains real with non-vanishing FD. 
With increasing $g$, part of the eigenspectrum forms a closed loop on the complex plane under the PBC [see Fig.~\ref{fig:2}(a2)], in contrast to the OBC case, where the eigenspectrum remains on the real axis [see Fig.~\ref{fig:2}(b2)].
The latter derives from the similarity transformation $\hat{H}_1(g,0)=S^{-1}\hat{H}_1(0,0)S$ in the basis ${(\hat{b}_1,\cdots,\hat{b}_{2M})}^T$ with $S=\text{diag}(e^g,\cdots,e^{2Mg})$, which leaves the eigenvalues unchanged for different values of $g$.
Crucially, while this transformation maps all eigenstates to boundary-localized skin modes under the OBC (under a finite $g$), it does not apply under the PBC. 
It follows that, if any bulk localized states are present under the PBC, their localization properties must be different under the OBC.

These boundary-sensitive localized states indeed exist at appropriately chosen $g$, illustrated in
Fig.~\ref{fig:2}(a2) as the eigenstates on the spectral branch within the loop with real eigenvalues, their FDs approaching unity.
In particular, for three randomly chosen eigenstates (marked by colored dots), their spatial distributions show sharp peaks near different localization centers in the bulk [see Fig.~\ref{fig:2}(c)] under the PBC. 
Meanwhile, all eigenstates of the system under the OBC are boundary-localized [see Fig.~\ref{fig:2}(d)].

The emergence of the boundary-sensitive localization can be understood and engineered through the Lyapunov exponents. 
While their sign pattern encodes the localization behavior of eigenstates with eigenenergy $E$ under different boundary conditions, the Lyapunov exponents 
at $g=0$ is related to those at finite $g$ through $\gamma_i(E) \rightarrow \gamma_i(E)-g $ (see Appendix~\ref{app:shift}), which offers a convenient way to adjust the sign pattern.

For instance, in the Hermitian case [Fig.~\ref{fig:2}(a1,b1)], the sign pattern is $(-,0,0,+)$ (listed as signs from $\gamma_1$ to $\gamma_4$) for all eigenstates, indicating extended or critical states under either boundary condition. 
With $g>0$, the sign pattern is shifted toward negative values. 
Importantly, for eigenstates on the central branch in Fig.~\ref{fig:2}(a2), the sign pattern is $(-,-,-,+)$.
For these states, the dominant Lyapunov exponents are $\gamma_3$ and $\gamma_4$ under the PBC, and their opposite signs suggest bulk localization. However, under the OBC, the dominant Lyapunov exponents are $\gamma_2$ and $\gamma_3$, which possess the same sign, indicating boundary localization typical of the skin modes.
On the other hand, for states on the outer loop in Fig.~\ref{fig:2}(a2), their corresponding pattern is $(-,-,0,+)$, consistent with their extended nature. 
Note that the eigenenergies of these states are dramatically changed under the OBC [see Fig.~\ref{fig:2}(b2)], wherein their sign pattern becomes $(-,-,-,+)$, indicating boundary localization.

\begin{figure}
\centering

\includegraphics[scale=0.19]{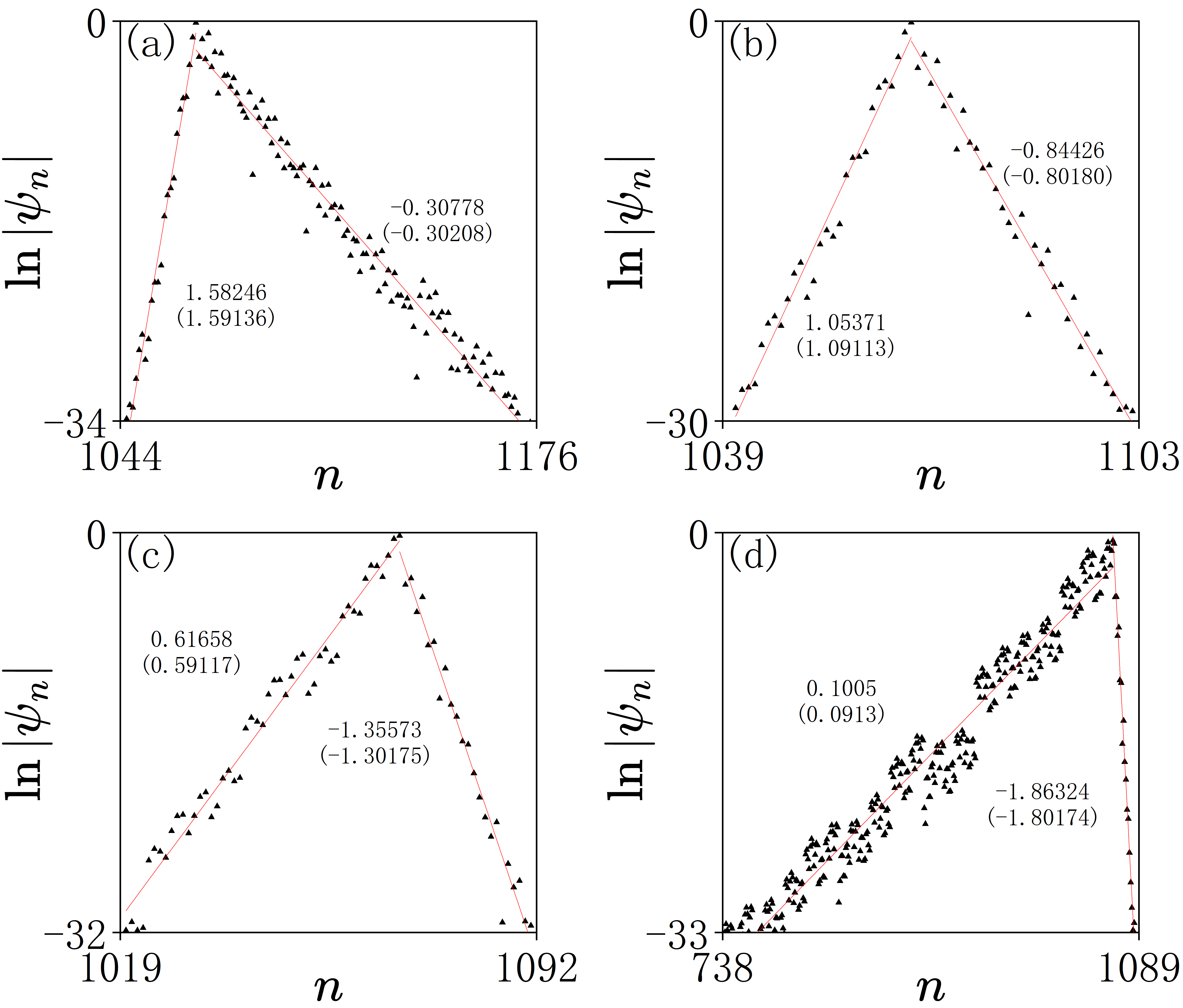}

\caption{Spatial distributions for eigenstates of $\hat{H}_{1}(g,0)$, with
$E=0.025$, and (a) $g=0.5$, (b) $g=1$, (c) $g=1.5$,
(d) $g=2$. Black dots represent numerical results, and red lines denote linear fits. The numbers outside the brackets represent the localization strengths obtained from the linear fits, and those within the brackets represent the corresponding $\gamma_{3}(E)$ and $\gamma_{4}(E)$. For all calculations, the system size is $N=10946$ and the PBC is imposed. \label{fig:3}}

\end{figure}

To further confirm the understanding above, we show in Fig.~\ref{fig:3} the spatial distribution of eigenstates with a fixed eigenenergy but under different $g$. In all cases, sharp peaks are observed near the localization centers, with the localization lengths on the two sides matching $\gamma_{3}(E)$ and $\gamma_{4}(E)$, respectively.

As such, these boundary-sensitive localized states necessarily exist for nonvanishing sets of Lyapunov exponents with an asymmetric sign pattern, that is, with an unequal number of positive and negative exponents.
For this to happen, three key ingredients are required:
(i) quasiperiodicity, which enables the emergence of localization;
(ii) non-Hermitian skin effect, which breaks the symmetry of the sign pattern; and
(iii) the presence of multiple Lyapunov exponents ($\geq 4$), which 
occurs for models with long-range hoppings or by introducing additional on-site degrees of freedom.

We note that boundary-sensitive localization has recently been reported in Ref.~\cite{SZLi2025}, termed as ``Anderson-skin dualism''. The ladder configuration therein, when viewed as a single-chain model with internal degrees of freedom, should give rise to four Lyapunov exponents. In the Appendix,, we show that the boundary-sensitive localized states therein can be naturally explained by our framework.

Physically, the boundary-sensitive localization under the PBC is akin to the inner skin effect in Ref.~\cite{sourav2023}, where the incompatibility of the PBC and quasiperiodicity effectively generates inner boundaries in the bulk, which becomes the various localization centers due to the non-Hermitian boundary sensitiveity.
Whereas under the OBC, nonreciprocal hoppings make eigenstates localize at the real physical boundaries.
These considerations suggest that 
boundary-sensitive localization should be a common phenomenon in non-Hermitian quasiperiodic systems with long-range hopping.

\section{Breakdown of the extended--localized duality relation}
The interplay of non-Hermiticity and quasiperiodicity also has interesting implications for the nonlocal state under the PBC.
To see this, we examine the dual model of $H_1(g,h)$ (see Appendix~\ref{app:duality})
\begin{flalign}
&\hat{H}_{2}(g,h)=\nonumber \\
&\sum_{m}(e^{-h}\hat{b}_{m}^{\dagger}\hat{b}_{m+1}+e^{h}\hat{b}_{m+1}^{\dagger}\hat{b}_{m})\sum_{s=1}^{2}\frac{\cos[\tau s\pi(2m+1)+isg]}{s^{3}},\label{eq:H2}
\end{flalign}
which features only nearest-neighbor hopping ($M=1$) but two quasiperiodic modulations parameterized by $s$.
From a numerical perspective, while the duality strictly holds in the thermodynamic limit for irrational $\tau$, it exactly holds in finite systems by imposing PBC with system size $N=F_j$ and adopting the rational approximation $\tau_{\rm RA}=F_{j-1}/F_j$, where $F_j$ is the $j$th Fibonacci number (see Appendix~\ref{app:duality}).
By design, both $H_1$ and $H_2$ possess the non-Hermitian skin effect under finite $g$ and $h$.

\begin{figure}
\centering

\includegraphics[scale=0.35]{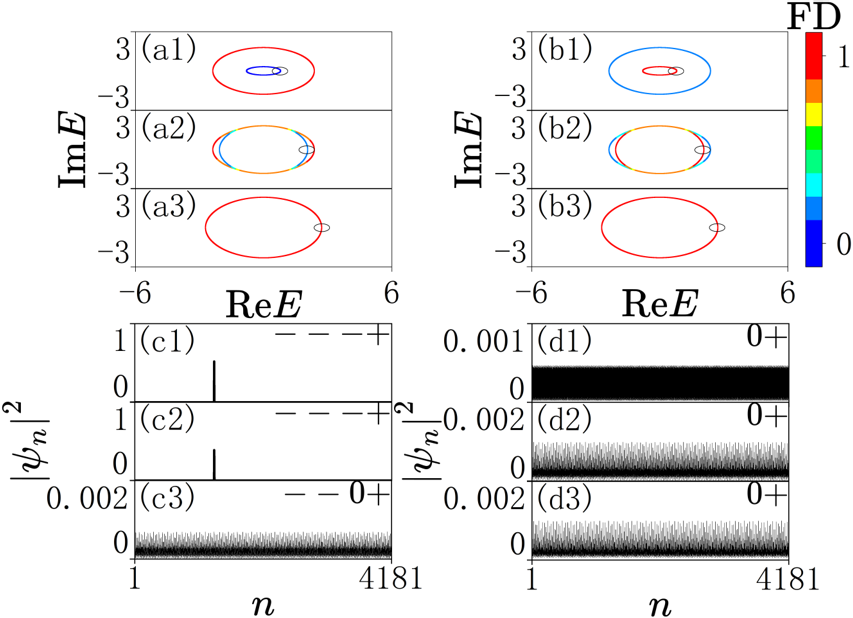}

\caption{
(a)(b) Eigenspectra of (a)
$\hat{H}_{1}(1,h)$  and (b) $\hat{H}_{2}(1,h)$, respectively, under the PBC, 
with color contours indicating the FD of the corresponding eigenstates. 
The parameters are (a1)(b1) $h=0.1$, (a2)(b2) $h=0.4$, and (a3)(b3) $h=0.6$. 
(c1-c3)(d1-d3) Spatial distributions of the eigenstates marked by
circles on the real axis in (a1–a3) and (b1–b3), respectively. 
The corresponding sign patterns of $\gamma_{i}(E)$ are indicated in the 
upper-right corners of (c)(d). 
For all calculations, the system size is fixed at $N=4181$ under the PBC, with the irrational number $\tau$ is approximated by $\tau_{\rm RA}=2584/4181$.
\label{fig:4}}
\end{figure}

According to the conventional extended--localized duality relation, an extended (localized) state in $H_1(g,h)$ should be mapped to a localized (extended) state in $H_2(g,h)$, whereas critical states remain critically localized in both models.
However, this duality relation does not always hold. 
In Fig.~\ref{fig:4}(a)(b), we show the PBC eigenspectra of both models, respectively, with a fixed $g$ and varying $h$.
When $h$ is small, the eigenvalues in the complex plane form two concentric rings. Under $\hat{H}_{1}$, the inner (outer) ring corresponds to localized (extended) states, while for $\hat{H}_{2}$ the situation is reversed. In this regime, the duality relation holds well.
As $h$ increases, the two rings start to merge, and the extended--localized duality relation breaks down in the overlapping region, although it still holds in the remaining parts of the spectrum.
In Fig.~\ref{fig:4}(a3)(b3), the rings completely overlap, and all eigenstates of both models are extended (with their FD approaching $1$), violating the aforementioned duality relation.
Note that no eigenstate degeneracy exists in the overlapping region, thus ruling out the possibility of a trivial hybridization of the extended and localized.

To confirm the observations above, we show typical spatial profiles of eigenstates of $\hat{H}_{1}$ in Fig.~\ref{fig:4}(c), and those of $\hat{H}_{2}$ but with the same eigenenergies in Fig.~\ref{fig:4}(d). 
We also compute the corresponding sign patterns of the Lyapunov exponents. 
In Fig.~\ref{fig:4}(c1)(c2) and (d1)(d2), the duality relation still holds, and the sign patterns are 
$(-,-,-,+)$ and $(0,+)$, consistent with the extended-localized duality relation.
By contrast, the sign patterns become $(-,-,0,+)$ and $(0,+)$ in Fig.~\ref{fig:4}(c3)(d3), as both eigenstates are extended. 
This conclusion is supported by finite-size-scaling analysis in Appendix~\ref{app:scaling}, which confirms the breakdown of the extended--localized duality relation in the thermodynamic limit.

The breakdown of extended--localized duality is counterintuitive, but not unreasonable. 
Since the dual transformation is essentially a Fourier transformation, the mapping from localized states to extended states can be understood as a manifestation of the Heisenberg–Pauli–Weyl uncertainty principle~\cite{ciatti2007}. However, the Fourier transform of an extended state can remain extended without violating this principle. 
 For a conventional extended wave function with a single momentum component, $\psi(x)\sim e^{ikx}$, its Fourier transform is localized in the momentum space, namely $\tilde{\psi}(k)\sim \delta(k)$, which follows the conventional extended--localized duality relation. However, for a more general extended state composed of multiple momentum components,
$
\psi(x)=\sum_{i} c_i e^{ik_i x},
$
its Fourier transform becomes
$
\tilde{\psi}(k)=\sum_i c_i \delta(k-k_i),$
which can itself remain extended in the momentum space due to the presence of multiple finite-weight momentum components.
From this viewpoint, the conventional extended--localized duality should not be regarded as universally guaranteed, but rather as a special consequence of simple single-frequency wave functions.

Note that the observed breakdown of extended--localized duality always occurs when both $\hat{H}_1$ and $\hat{H}_2$ support vanishing Lyapunov exponents and 
possess nonlocal eigenstates at the same energy $E$.
This is facilitated by the non-Hermtian skin effect, which breaks the symmetry in the sign pattern of the Lyapunov exponents, enabling vanishing Lyapunov exponents in both models at the same energy. Thus, the interplay between nonreciprocal hopping and quasiperiodic modulation with imaginary phases provides a universal non-Hermitian mechanism for realizing extended--extended duality, a scenario absent in conventional Hermitian settings (see Appendix~\ref{app:dualitybreak}).

\section{Summary}
From the perspective of Lyapunov exponents,
we demonstrate insight and control of anomalous boundary-sensitive localization and the breakdown of the extended--localized duality relation in non-Hermitian quasiperiodic lattices,
and show that these phenomena are generic in non-Hermitian quasiperiodic systems with long-range hopping.
While our models can be implemented on existing quantum simulation platforms such as the Rydberg atoms~\cite{wang2025}, our work highlights the dramatic impact of non-Hermiticity on quasiperiodic systems.

\begin{acknowledgments}
This work is supported by the National Natural Science Foundation of China (Grant No. 12374479), and the Innovation Program for Quantum Science and Technology (Grant No. 2021ZD0301904). T.L. is supported by National Natural Science Foundation
of China (Grant No.12504189), and the Guangdong Basic and Applied Basic Research Foundation (Grant No.2026A1515012098).
\end{acknowledgments}

\appendix

\renewcommand{\thesection}{\Alph{section}}
\renewcommand{\thefigure}{\Alph{section}\arabic{figure}}
\renewcommand{\thetable}{\Alph{section}\Roman{table}}
\setcounter{figure}{0}
\renewcommand{\theequation}{\Alph{section}\arabic{equation}}
\setcounter{equation}{0}

\section{Derivation of the transfer matrix}\label{app:transfer}
Starting from the eigenvalue equations of $\hat{H}$ in the main text
\begin{equation}
\sum_{t=-M}^{M}J_{n,n+t}\psi_{n+t}=E\psi_{n}, \quad \quad  n = 1,2,...N, \label{eq:eigen_equa_supp}
\end{equation}
one can rewrite these eigenvalue equations in terms of a $2M \times 2M$ transfer matrices

\begin{equation}
\begin{pmatrix}\psi_{n+M}\\
\psi_{n+M-1}\\
\vdots\\
\psi_{n-M+1}
\end{pmatrix}=T_{n}\begin{pmatrix}\psi_{n+M-1}\\
\psi_{n+M-2}\\
\vdots\\
\psi_{n-M}
\end{pmatrix},\label{eq:trans_def_supp}
\end{equation}
with
\begin{widetext}
\begin{equation}
T_{n}(E)=\begin{pmatrix}-\frac{J_{n,n+M-1}}{J_{n,n+M}} & -\frac{J_{n,n+M-1}}{J_{n,n+M}} & \cdots & -\frac{J_{n,n+1}}{J_{n,n+M}} & \frac{E-J_{n,n}}{J_{n,n+M}} & -\frac{J_{n,n-1}}{J_{n,n+M}} & \cdots & -\frac{J_{n,n-M}}{J_{n,n+M}}\\
1\\
 & 1\\
 &  & \ddots\\
 &  &  & 1\\
 &  &  &  & 1\\
 &  &  &  &  & 1\\
 &  &  &  &  &  & \ddots
\end{pmatrix}, n = 1,2,...,N.\label{eq:trans_form_supp}
\end{equation}
\end{widetext}
Note that this matrix has nonzero elements only on the first row and the subdiagonal, and propagates the wave function by one lattice site at each step.

By introducing the supercell formalism and rearranging the eigenvalue equations in Eq.~(\ref{eq:eigen_equa_supp}), one can obtain alternatively~\cite{Sun2025},
\begin{equation}
\begin{pmatrix}\tilde{\Psi}_{n+1}\\
\tilde{\Psi}_{n}
\end{pmatrix}=\tilde{T}_{n}\begin{pmatrix}\tilde{\Psi}_{n}\\
\tilde{\Psi}_{n-1}
\end{pmatrix},\label{eq:trans_tilde_supp}
\end{equation}
where $\tilde{\Psi}_{n}=(\psi_{nM},\psi_{nM-1},\cdots,\psi_{(n-1)M+1})^{T}$ denotes the supercell wave function.

The supercell transfer matrix $\tilde{T}_{n}$ propagates the wave function by one supercell at each step, in contrast to $T_n(E)$. 
The two transfer matrices are related through
\begin{equation}
\tilde{T}_{n}=\prod_{m=1}^{M}T_{(n-1)M+m}.
\end{equation}
It should be noted that these two transfer matrices yield the same Lyapunov exponents.

\section{Nonreciprocity-Induced Shift of Lyapunov Exponents}\label{app:shift}

When the parameter $g$ in $\hat{H}_1(g,h)$ of the main text is switched on from $0$ to a nonzero value, the hopping coefficients $J_{n,m}(g)$ in Eq.~(\ref{eq:eigen_equa_supp}) are modified as
\begin{equation}
J_{n,m}(g)= e^{(m-n)g}J_{n,m}(g=0).
\label{eq:J_g_supp}
\end{equation}
This induces the following transformation of the single-site transfer matrix:
\begin{equation}
T_n(g,E)= e^{-g} \, S \, T_n(g=0,E) \, S^{-1},
\label{eq:Tn_g_supp}
\end{equation}
where $S=\mathrm{diag}(e^g,e^{2g},\dots,e^{2Mg})$.

Accordingly, the full transfer matrix becomes
\begin{widetext}
\begin{equation}
T(g,E)=\prod_{n=1}^{N}T_n(g,E)
=
e^{-Ng} \, S \, \prod_{n=1}^{N}T_n(g=0,E) \, S^{-1}
=
e^{-Ng} \, S \, T(g=0,E) \, S^{-1}.
\label{eq:T_g_supp}
\end{equation}

It then follows that
\begin{equation}
\frac{1}{2N}\ln [T^\dagger(g,E)T(g,E)]
=
\frac{1}{2N}\ln [S T^\dagger(g=0,E) T(g=0,E) S^{-1}] - g .
\label{eq:T_posi_g_supp}
\end{equation}
\end{widetext}

Since the Lyapunov exponents $\gamma_i(E)$ are given by the eigenvalues of $\frac{1}{2N}\ln [T^\dagger(E)T(E)]$ in the limit $N\rightarrow +\infty$, and similarity transformations do not change eigenvalues, the Lyapunov exponents are shifted according to
\begin{equation}
\gamma_i(g,E)=\gamma_i(g=0,E)-g, \quad \quad i=1,2,...,2M.
\label{eq:gamma_g_supp}
\end{equation}

\section{Duality between $\hat{H}_{1}$ and $\hat{H}_{2}$}\label{app:duality}

In this section, we demonstrate the duality between $\hat{H}_{1}$ and $\hat{H}_{2}$ in the main text. For completeness, we rewrite them here:
\begin{widetext}
\begin{equation}
\hat{H}_{1}(g,h)=\sum_{1\leq|m-n|\leq2}e^{(n-m)g}\hat{b}_{m}^{\dagger}\hat{b}_{n}\frac{\cos[\tau\pi(m+n)+ih]}{|m-n|^{a}},\label{eq:H1_supp}
\end{equation}
and
\begin{equation}
\hat{H}_{2}(g,h)=\sum_{m}(e^{-h}\hat{b}_{m}^{\dagger}\hat{b}_{m+1}+e^{h}\hat{b}_{m+1}^{\dagger}\hat{b}_{m})\sum_{s=1}^{2}\frac{\cos[\tau s\pi(2m+1)+isg]}{s^{a}}.\label{eq:H2_supp}
\end{equation}
\end{widetext}

Applying the dual transformation
\begin{equation}
\hat{b}_{n} = \frac{1}{\sqrt{N}} \sum_{k=1}^{N} e^{-2i\pi \tau nk} \hat{b}_{k}, \quad n=1,2,\dots,N,
\label{eq:dual_trans_supp}
\end{equation}
we obtain
\begin{widetext}
\begin{flalign}
\hat{H}_{1}(g,h) 
= & \sum_{1\leq|m-n|\leq2} e^{(n-m)g} \hat{b}_{m}^{\dagger}\hat{b}_{n} \frac{\cos[\tau\pi(m+n)+ih]}{|m-n|^{a}} \nonumber \\
= & \sum_{t=\pm1,\pm2} \sum_{n} e^{-tg} \hat{b}_{n+t}^{\dagger} \hat{b}_{n} \frac{\cos[\tau\pi(2n+t)+ih]}{|t|^{a}} \nonumber \\
= & \frac{1}{2N} \sum_{t=\pm1,\pm2} \frac{e^{-tg}}{|t|^{a}} \sum_{nkl} \hat{b}_{k}^{\dagger} \hat{b}_{l} \, e^{2i\pi \tau (n+t)k} e^{-2i\pi \tau n l} \big( e^{i \tau \pi (2n+t)} e^{-h} + e^{-i \tau \pi (2n+t)} e^{h} \big) \nonumber \\
= & \frac{1}{2N} \sum_{t=\pm1,\pm2} \frac{e^{-tg}}{|t|^{a}} \sum_{nkl} \hat{b}_{k}^{\dagger} \hat{b}_{l} \big( e^{-h} e^{2\pi i \tau n (k-l+1)} e^{\pi i \tau t (2k+1)} + e^{h} e^{2\pi i \tau n (k-l-1)} e^{\pi i \tau t (2k-1)} \big) \nonumber \\
= & \frac{1}{2} \sum_{t=\pm1,\pm2} \frac{e^{-tg}}{|t|^{a}} \sum_{kl} \Big( e^{-h} e^{\pi i \tau t (2k+1)} \frac{1}{N} \sum_{n} \hat{b}_{k}^{\dagger} \hat{b}_{l} e^{2\pi i \tau n (k-l+1)} \nonumber \\
& + e^{h} e^{\pi i \tau t (2k-1)} \frac{1}{N} \sum_{n} \hat{b}_{k}^{\dagger} \hat{b}_{l} e^{2\pi i \tau n (k-l-1)} \Big) \nonumber \\
= & \frac{1}{2} \sum_{t=\pm1,\pm2} \frac{e^{-tg}}{|t|^{a}} \sum_{kl} \big( e^{-h} e^{\pi i \tau t (2k+1)} \hat{b}_{k}^{\dagger} \hat{b}_{l} \, \delta_{k,l-1} + e^{h} e^{\pi i \tau t (2k-1)} \hat{b}_{k}^{\dagger} \hat{b}_{l} \, \delta_{k,l+1} \big) \nonumber \\
= & \frac{1}{2} \sum_{t=\pm1,\pm2} \frac{e^{-tg}}{|t|^{a}} \sum_{l} \big( e^{-h} e^{\pi i \tau t (2l+1)} \hat{b}_{l}^{\dagger} \hat{b}_{l+1} + e^{h} e^{\pi i \tau t (2l+1)} \hat{b}_{l+1}^{\dagger} \hat{b}_{l} \big) \nonumber \\
= & \sum_{l} \big( e^{-h} \hat{b}_{l}^{\dagger} \hat{b}_{l+1} + e^{h} \hat{b}_{l+1}^{\dagger} \hat{b}_{l} \big) \sum_{t=\pm1,\pm2} \frac{e^{-tg} e^{\pi i \tau t (2l+1)}}{2 |t|^{a}} \nonumber \\
= & \sum_{l} \big( e^{-h} \hat{b}_{l}^{\dagger} \hat{b}_{l+1} + e^{h} \hat{b}_{l+1}^{\dagger} \hat{b}_{l} \big) \sum_{t=1}^{2} \frac{\cos[\tau t \pi (2l+1) + i t g]}{t^{a}}, 
\label{eq:H1_to_H2}
\end{flalign}
\end{widetext}
which reproduces exactly $\hat{H}_{2}(g,h)$ in the dual basis, thus confirming the duality between $H_1(g,h)$ and $H_2(g,h)$. Here, $\delta_{k,l}$ is the Kronecker delta. In the derivation, we have used the identity
\begin{equation}
\lim_{N \rightarrow +\infty} \frac{1}{N} \sum_{n=1}^{N} e^{2\pi i \tau n (k-l)} = \delta_{k,l},
\label{eq:id_supp}
\end{equation}
which holds strictly for irrational $\tau$.

While the above derivation strictly holds in the thermodynamic limit and for irrational $\tau$, the duality relation can also be realized for finite systems under appropriate conditions.

By imposing periodic boundary conditions $\hat{b}_{n+N} = \hat{b}_{n}$ for a finite system size $N = F_j$ and adopting the rational approximation $\tau_{\rm RA} = F_{j-1}/F_j$, where $F_j$ is the $j$-th Fibonacci number defined recursively as $F_j = F_{j-1} + F_{j-2}$ with $F_1 = 0$ and $F_2 = 1$, the derivation above remains valid.

Specifically, under the dual transformation
\begin{equation}
\hat{b}_{n} = \frac{1}{\sqrt{N}} \sum_{k=1}^{N} e^{-2 i \pi \tau_{\rm RA} n k} \hat{b}_{k}, \quad n=1,2,\dots,N,
\label{eq:dual_trans_ra_supp}
\end{equation}
and using the identity
\begin{equation}
\frac{1}{N} \sum_{n=1}^{N} e^{2 \pi i \tau_{\rm RA} n (k-l)} = \delta_{k,l},
\label{eq:id_ra_supp}
\end{equation}
the derivation in Eq.~(\ref{eq:H1_to_H2}) still holds. 
Such a construction of duality in finite-size systems facilitate numerical calculations.

\section{Scaling analysis of the boundary-sensitive localized states and dual extended states}\label{app:scaling}

\begin{figure}
\centering
\includegraphics[scale=0.2]{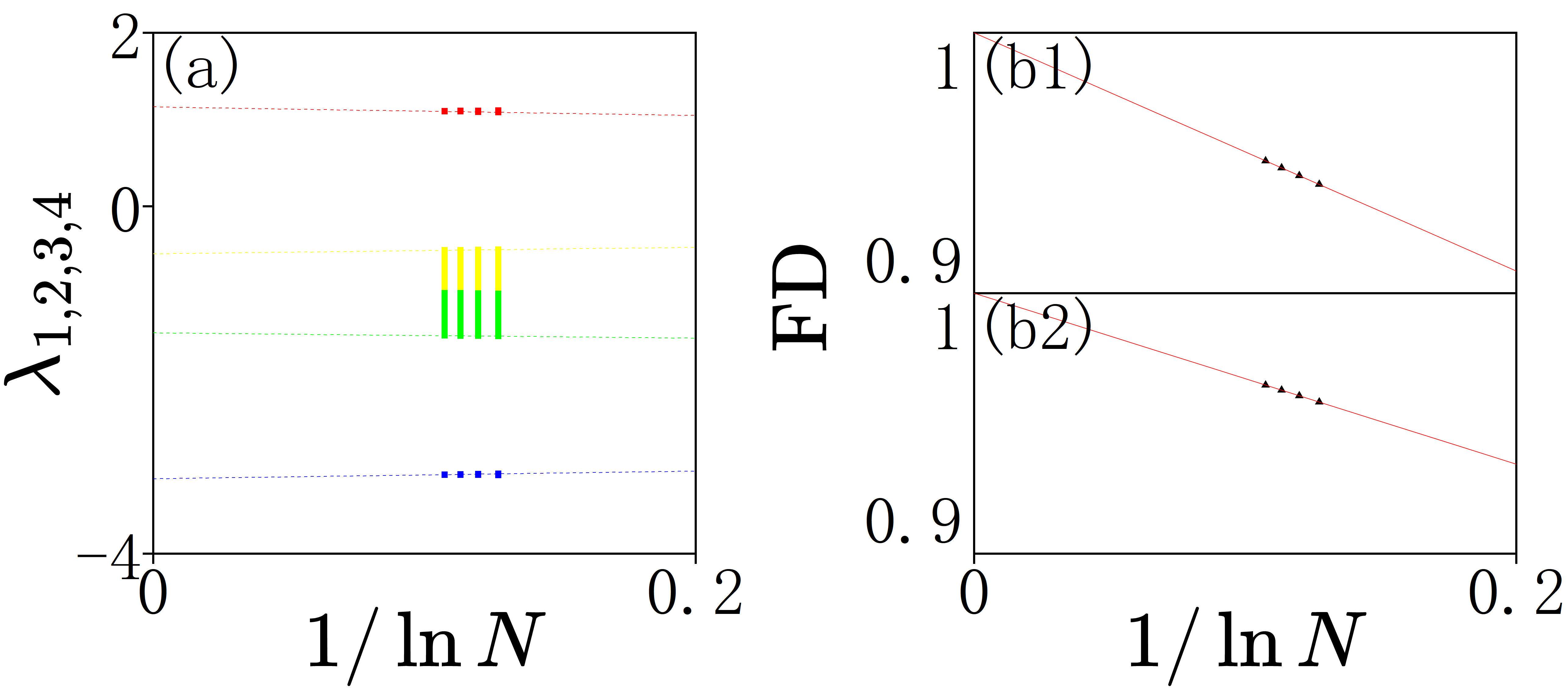}
\caption{
(a) Lyapunov exponents of the boundary-sensitive localized states in Fig.~2(a3) of the main text, plotted as functions of the inverse logarithm of the system size, $1/\ln N$. Points of different colors correspond to $\gamma_1$ (red), $\gamma_2$ (yellow), $\gamma_3$ (green), and $\gamma_4$ (blue), respectively. Dashed lines show linear fittings for the smallest $\gamma_1$, the largest $\gamma_2$, the smallest $\gamma_3$, and the largest $\gamma_4$ for each $N$. The sign pattern $(---+)$ of the Lyapunov exponents remains unchanged with increasing system size.  
(b) Scaling analysis of the FD for a dual pair of eigenstates of the models $H_1$ and $H_2$, marked by circles on the real axis in Fig.~4(a3) and (b3) of the main text. Black dots are numerical results, and red lines are linear fittings. Both FDs approach $1$ as $N$ approaches inifinity, confirming that these states are extended.
\label{fig1}}
\end{figure}

In this section, we perform a scaling analysis of the boundary-sensitive localized states in Fig.~2(a3) and the dual extended states in Fig.~4(a3) and (b3) of the main text. As shown in Fig.~\ref{fig1}(a), the sign pattern $(---+)$ of the Lyapunov exponents for the boundary-sensitive localized states remains unchanged with increasing system size, indicating that these localized states are not artifacts from finite-size effects. Figure~\ref{fig1}(b) shows the scaling of the fractal dimensions (FDs) of two dual extended eigenstates of $H_1$ and $H_2$. Both FDs approach $1$ as $N$ increases, confirming that these states are indeed extended in the thermodynamic limit.

\section{More on the breakdown of extended--localized duality}\label{app:dualitybreak}
In this section, we provide a more detailed discussion on the breakdown of extended--localized duality.

 We now consider the following nearest-neighbor hopping Hamiltonians
\begin{align}
\hat{H}_1=\sum_n(e^g\hat{a}_n^\dagger\hat{a}_{n+1}+e^{-g}\hat{a}_{n+1}^\dagger\hat{a}_n)\cos(\pi\tau(2n+1)+ih),\\
\hat{H}_2=\sum_n(e^{-h}\hat{a}_n^\dagger\hat{a}_{n+1}+e^h\hat{a}_{n+1}^\dagger\hat{a}_n)\cos(\pi\tau(2n+1)+ig).
\end{align}
These models, being dual to each other, are studied in Ref.~\cite{Cai2025}, but not in the context of duality breaking.
When either $g=0$ or $h=0$, all eigenstates of both models are critically localized~\cite{Cai2022,Cai2025}, consistent with conventional duality relations (critical--critical). However, when $g\neq0$ and $h\neq0$, both models exhibit extended eigenstates simultaneously~\cite{Cai2025}, signaling the breakdown of the conventional extended--localized duality.

The physical origin of this phenomenon lies in the combined effect of the imaginary phases ($h$ in $H_1$ and $g$ in $H_2$) and the nonreciprocal hopping. For $\hat{H}_1,\hat{H}_2$ with zero imaginary phases, there are inhomogeneously distributed zeros (IDZs) in the quasiperiodic hopping terms, which are believed to lead to the emergence of critical states~\cite{Qin2026, Zhou2023}. However, a finite imaginary phase eliminates IDZs of the quasiperiodic hopping terms, which facilitates the unidirectional flow in the bulk. Consequently, when the relevant Lyapunov exponents vanish in both models and the corresponding eigenstates become nonlocal, the eigenstates tend to be extended rather than critically localized.

Such a picture is further clarified by the extreme case with
\begin{widetext}
\begin{align}
\hat{H}_1=\sum_n(e^{g-h}e^{i(2n+1)\pi\alpha}+e^{g+h}e^{-i(2n+1)\pi\alpha})|n\rangle\langle n+1| + (e^{-g-h}e^{i(2n+1)\pi\alpha}+e^{-g+h}e^{-i(2n+1)\pi\alpha})|n+1\rangle\langle n|,\\
\hat{H}_2=\sum_n(e^{-h-g}e^{i(2n+1)\pi\alpha}+e^{-h+g}e^{-i(2n+1)\pi\alpha})|n\rangle\langle n+1| + (e^{h-g}e^{i(2n+1)\pi\alpha}+e^{h+g}e^{-i(2n+1)\pi\alpha})|n+1\rangle\langle n|.
\end{align}
\end{widetext}

In the limit $g, h\gg0$, these Hamiltonians reduce to
\begin{align}
\hat{H}_1=e^{g+h}\sum_ne^{-i(2n+1)\pi\alpha}|n\rangle\langle n+1|,\\
\hat{H}_2=e^{g+h}\sum_ne^{-i(2n+1)\pi\alpha}|n+1\rangle\langle n|,
\end{align}
where none of the hopping terms vanish. Apart from the overall factor $e^{g+h}$, both reduced Hamiltonians are equivalent to unitary hopping matrices and can be directly verified to exhibit extended eigenstates. 

As such, the interplay between nonreciprocal hopping and quasiperiodic modulation with imaginary phases provides a universal non-Hermitian mechanism for the extended--extended duality. This mechanism is fundamentally different from previously known Hermitian dualities, but is closely tied to the non-Hermitian skin effect and tunable through the LE patterns.

\section{Boundary-sensitive localized states in Ref.~\cite{SZLi2025}}\label{app:szli}

In this section, we discuss the boundary-sensitive localized states reported in Ref.~\cite{SZLi2025} and show that they can be naturally understood within our transfer-matrix framework.

We begin with the reduced model in Ref.~\cite{SZLi2025} [Eq.~(5) therein]
\begin{equation}
J_R \psi_{a,n-1} + \frac{t^2}{E-V_n} \psi_{a,n} + J_L \psi_{a,n+1} = E\psi_{a,n}. \label{eq:reduced}
\end{equation}
where $V_n=2\lambda \cos(2\pi\tau n)$, $J_L=e^g$, $J_R=e^{-g}$, and $\tau=\frac{\sqrt{5}-1}{2}$. In the Hermitian case $g=0$, Eq.~\eqref{eq:reduced} yields a pair of Lyapunov exponents $\pm \gamma$ with $\gamma>0$ \begin{equation}
\gamma(E) = \begin{cases} 
\ln \left| \frac{\lambda (E + \sqrt{E^2 - 4})}{(E + \sqrt{E^2 - 4\lambda^2})} \right|, & |E| \ge 2\lambda \ \& \ |E|   \ge 2, \\ 
\ln \left| \frac{E + \sqrt{E^2 - 4}}{2} \right|, & |E| < 2\lambda \ \& \ |E| \ge 2, \\ 
\ln \left| \frac{2\lambda}{(E + \sqrt{E^2 - 4\lambda^2})} \right|, & |E| \ge 2\lambda \ \& \ |E| < 2, \\ 
0, & |E| < 2\lambda \ \& \ |E| < 2.
\end{cases} \label{eq:gamma}
\end{equation}

Next consider the non-Hermitian case $g\neq 0$. The Lyapunov exponents of Eq.~\eqref{eq:reduced} are shifted to
$\gamma_1(E)=-\gamma(E)-g$ and $\gamma_2(E)=\gamma(E)-g$.
For the boundary-sensitive localized states shown in Fig.~2 of Ref.~\cite{SZLi2025}, we have $|E| < 2\lambda$ and $|E| < 2$, hence $\gamma(E)=0$. Therefore, the signs of the Lyapunov exponents are $(-,-)$ for $g>0$, and $(+,+)$ for $g<0$.

Since the ladder configuration in Ref.~\cite{SZLi2025} can be equivalently mapped to a single-chain model with next-nearest-neighbor (NNN) hopping ($M=2$), the corresponding transfer matrix is $4\times 4$, implying four Lyapunov exponents.
We therefore consider adding the NNN hopping terms to the reduced model
\begin{widetext}
\begin{equation}
t_2 \psi_{a,n-2}+J_R \psi_{a,n-1} + \frac{t^2}{E-V_n} \psi_{a,n} + J_L \psi_{a,n+1}+t_2 \psi_{a,n+2}= E\psi_{a,n}.\label{eq:NNN}
\end{equation}
\end{widetext}
Note that taking the limit $t_2 \rightarrow 0$ recovers Eq.~(\ref{eq:reduced}), but two additional Lyapunov exponents generally exist. In the following, we label the two models as the NN and NNN model, respectively.

We employ Oseledec's Multiplicative Ergodic Theorem (MET) of the full transfer matrix $M_N = \prod_{n=1}^N T_n$ of the NNN model, where
\begin{equation}
T_n = \begin{pmatrix} -\frac{J_L}{t_2} & -\frac{\tilde{V}_n}{t_2} & -\frac{J_R}{t_2} & -1 \\ 1 & 0 & 0 & 0 \\ 0 & 1 & 0 & 0 \\ 0 & 0 & 1 & 0 \end{pmatrix},\label{eq:T_n_NNN}
\end{equation}
with $\tilde{V}_n(E)=\frac{t^2}{E-V_n}-E$.
According to MET, the maximum Lyapunov exponent $\gamma_m$ is strictly defined by the asymptotic growth rate of the operator norm
\begin{equation}
\gamma_m = \lim_{N \to \infty} \frac{1}{N} \ln \| M_N \|_2, \label{eq:gamma_norm}
\end{equation}
where $\| M_N \|_2$ is the matrix spectral norm (the 2-norm) of $M_N$, defined as the maximum stretching factor of $M_N$ acting on a vector space
\begin{equation}
\| M_N \|_2 = \sup_{\|\mathbf{x}\|_2 = 1} \|M_N\mathbf{x}\|_2, \label{eq:2_norm}
\end{equation}
where $\|\mathbf{x}\|_2=\sqrt{|x_1|^2+|x_2|^2+|x_3|^2+|x_4|^2}$.

We factor out the parameter $t_2$ from the local $4 \times 4$ transfer matrix by writing $T_n = \frac{1}{t_2} A_n$, where the scaled matrix $A_n$ takes the form
\begin{equation}
A_n = \begin{pmatrix} 
-J_L & -\tilde{V}_n & -J_R & -t_2 \\ 
t_2 & 0 & 0 & 0 \\ 
0 & t_2 & 0 & 0 \\ 
0 & 0 & t_2 & 0 
\end{pmatrix}.\label{eq:A_n}
\end{equation}
Substituting this into the global product yields $M_N = t_2^{-N} \prod_{n=1}^N A_n$, then Eq.~(\ref{eq:gamma_norm}) can be decomposed as
\begin{equation}
\gamma_m = -\ln |t_2| + \tilde{\gamma},\label{eq:gamma_de}
\end{equation}
where $\tilde{\gamma} = \lim_{N \to \infty} \frac{1}{N} \ln \left\| \prod_{n=1}^N A_n \right\|_2$.  For $\tilde{\gamma}$, we consider the limit $t_2 \rightarrow 0$, for which

\begin{equation}
\prod_{n=1}^N A_n^{(t_2=0)} = \begin{pmatrix} 
(-J_L)^N & J_L^{N-1}\tilde{V}_1 & J_L^{N-1}J_R & 0 \\ 
0 & 0 & 0 & 0 \\ 
0 & 0 & 0 & 0 \\ 
0 & 0 & 0 & 0 
\end{pmatrix}. \label{eq:A_prod} 
\end{equation}
A fundamental property of the matrix spectral norm (the 2-norm) dictates that it is bounded from below by the absolute value of any individual matrix element. Applying this to the $(1,1)$ element of $\prod_{n=1}^N A_n^{(t_2=0)}$ in Eq.~(\ref{eq:A_prod}), we obtain the lower bound
\begin{equation}
\tilde{\gamma} \ge \lim_{N \to \infty} \frac{1}{N} \ln \left( |-J_L|^N \right) = \ln |J_L|.\label{eq:bound}
\end{equation}

Combining Eqs.~\eqref{eq:gamma_de} and \eqref{eq:bound}, we find that
\begin{equation}
\gamma_m \geq -\ln |t_2| + \ln |J_L|.
\end{equation}
As $t_2 \rightarrow 0$ , the term $-\ln |t_2|$ diverges to $+\infty$, which indicates the maximum Lyapunov exponent $\gamma_m$ diverges to $+\infty$ in the $NNN$ model.

Moreover, according to Eq.~(\ref{eq:T_n_NNN}), we have $\det(T_n)=1$, and 
\begin{equation}
\begin{aligned}
\sum_s \gamma_s 
&= \lim_{N \to \infty} \text{Tr} \left[ \frac{1}{2N} \ln (M_N^\dagger M_N) \right] \\
&= \lim_{N \to \infty} \frac{1}{2N} \ln \left[ \det (M_N^\dagger M_N) \right] \\
&= \lim_{N \to \infty} \frac{1}{N} \sum_{n=1}^N \ln |\det(T_n)| \\
&= 0,
\end{aligned}\label{eq:zero}
\end{equation}
where $M_N=\prod_{n=1}^N T_n$.
Hence, the presence of a diverging positive Lyapunov exponent implies the existence of a corresponding diverging negative Lyapunov exponent in the $t_2 \rightarrow 0$ limit.

In summary, adding NNN hoppings with coefficient $t_2 \rightarrow 0$ introduces a pair of additional Lyapunov exponents $\gamma_- \rightarrow -\infty$ and $\gamma_+ \rightarrow +\infty$. 
The full Lyapunov sign structure is therefore $(-,+,+,+)$ or $(-,-,-,+)$, which precisely corresponds to the boundary-sensitive localized states within our framework. Moreover, the negative (positive) diverging Lyapunov exponent reflects the extremely sharp localization associated with the $(-,+,+,+)$ ($(-,-,-,+)$) states, consistent with Fig.~2(f) of Ref.~\cite{SZLi2025} where the left side of the localization center are sharp.

In Ref.~\cite{SZLi2025}, the Lyapunov exponents of the NN model are defined as
\begin{equation}
\gamma_{L/R}=\max \{\gamma \pm g, 0 \},\label{eq:gamma_LR}
\end{equation}
where $\gamma$ is given by Eq.~\eqref{eq:gamma}. The quantities $\gamma_{L}$ and $\gamma_{R}$ are introduced therein to characterize the decay rates on the left- and right-hand sides of the localization center, respectively. By construction, they are constrained to be nonnegative. In contrast, no such constraint is imposed in our framework. Physically, different Lyapunov exponents $\gamma_i(E)$ correspond to different decay channels of an eigenstate with eigenenergy $E$. The contribution of the $i$th channel to the spatial profile of the eigenstate scales as $e^{\gamma_i(E)n}$. Consequently, channels with positive $\gamma_i$ correspond to exponentially growing amplitudes along the direction with increasing lattice index, whereas channels with negative $\gamma_i$ correspond to exponential decay.

For the boundary-sensitive localized states shown in Fig.~2 of Ref.~\cite{SZLi2025}, the quantities $\gamma_{L}$ and $\gamma_{R}$ are given by $(\gamma_L,\gamma_R)=(0,\gamma_R)$ with $g>0$.
While $\gamma_R$ correctly characterizes the decay on the right-hand side of the localization center, the sharp decay on the left side cannot be captured by $\gamma_L=0$. 
As discussed above, the left-hand side of the boundary-sensitive localized states shown in Fig.~2 of Ref.~\cite{SZLi2025} is associated with a Lyapunov exponent that diverges to $+\infty$ rather than vanishing. This indicates that the pair $(\gamma_L,\gamma_R)$ does not provide a complete characterization of these boundary-sensitive localized states.

Under our framework, however, the localization properties on both sides of the localization center are naturally encoded in the Lyapunov spectrum.
 Specifically, the transfer matrix yields four Lyapunov exponents, which we denote in ascending order by $\gamma_1<\gamma_2<\gamma_3<\gamma_4$, with the sign pattern $(-, -, -, +)$. By contrast, the $2\times2$ transfer matrix of the reduced model [Eq.~(F13)] captures only the middle pair, $\gamma_2$ and $\gamma_3$. For boundary-sensitive localized states, however, the localization profile is governed by the pair of Lyapunov exponents with opposite signs that are closest to zero. In the present case, these are $\gamma_3$ and $\gamma_4$, rather than $\gamma_2$ and $\gamma_3$. Consequently, the reduced description by Eq.~(R13) fails to capture the sharp decay on one side of the localization center. By retaining the full Lyapunov spectrum, our framework provides a complete characterization of the asymmetric localization profile of boundary-sensitive localized states.

\end{document}